\documentclass{aastex}
\usepackage{spr-astr-addons}
\usepackage{url}

\RequirePackage{color}

\begin{document}

\title{``Dark energy" in the Local Void}
\slugcomment{Not to appear in Nonlearned J., 45.}
%% Running heads
\shorttitle{``Dark energy" in the Local Void}
\shortauthors{M. Villata}

\author{M. Villata}
\affil{INAF, Osservatorio Astronomico di Torino, Via Osservatorio 20, I-10025 Pino Torinese (TO), Italy\\
%%%modificato
e-mail: villata@oato.inaf.it}

\begin{abstract}
The unexpected discovery of the accelerated cosmic expansion in 1998 has filled the Universe with the embarrassing presence of an unidentified ``dark energy", or cosmological constant, devoid of any physical meaning. While this standard cosmology seems to work well at the global level, improved knowledge of the kinematics and other properties of our extragalactic neighborhood indicates the need for a better theory. We investigate whether the recently suggested repulsive-gravity scenario can account for some of the features that are unexplained by the standard model. Through simple dynamical considerations, we find that the Local Void could host an amount of antimatter ($\sim5\times10^{15}\,M_\odot$) roughly equivalent to the mass of a typical supercluster, thus restoring the matter-antimatter symmetry. The antigravity field produced by this ``dark repulsor" can explain the anomalous motion of the Local Sheet away from the Local Void, as well as several other properties of nearby galaxies that seem to require void evacuation and structure formation much faster than expected from the standard model. At the global cosmological level, gravitational repulsion from antimatter hidden in voids can provide more than enough potential energy to drive both the cosmic expansion and its acceleration, with no need for an initial ``explosion" and dark energy. Moreover, the discrete distribution of these dark repulsors, in contrast to the uniformly permeating dark energy, can also explain dark flows and other recently observed excessive inhomogeneities and anisotropies of the Universe.
\end{abstract}

\keywords{Gravitation --- Cosmology: theory --- Dark energy --- Large-scale structure of Universe}

%\section*{}
%\label{sec:intro}

\section{Introduction}

Since the end of the last century, observations of high-redshift type Ia supernovae have unexpectedly shown that the cosmic expansion is currently in an acceleration phase \citep[e.g.][]{rie98,per99}, whose physical cause is unknown. Formally, this acceleration is ascribed to an additional term having a negative pressure in the expansion equations, which represents about 75\% of the total energy density of the Universe, in the simplest case corresponding to a cosmological constant, perhaps associated to the energy of the quantum vacuum. Besides this standard cosmology of the $\Lambda$CDM model, various alternatives have been proposed to explain the cosmic speed-up, invoking scalar fields or modifications of general relativity, such as extensions to extra dimensions or higher-order curvature terms \citep[e.g.][]{ame00,dva00,car04,cap05}. In a variant of these alternative theories, \citet{vil11} proposes to extend general relativity to antimatter, intended as CPT-transformed matter, whose immediate result is the prediction of a gravitational repulsion between matter and antimatter, which, with antimatter hidden in cosmic voids, could explain the accelerated expansion.

Knowledge of peculiar motions and spatial distribution of galaxies allows us to explore gravitational interactions within and among clusters. This is particularly true for our extragalactic neighborhood, where distances can be measured with higher precision. Through a study performed on a database of about 1800 galaxies within $3000\rm\,km\,s^{-1}$, \citet{tul08} find that the peculiar velocity of the Local Sheet decomposes into three dominant, almost orthogonal components. One of them (of $185\rm\,km\,s^{-1}$) is ascribed to the gravitational pull of the Virgo Cluster and its surroundings, while a second, larger component (of $455\rm\,km\,s^{-1}$) seems to be due to an attraction on larger scales nearly in the direction of the Centaurus Cluster. The third component ($259\rm\,km\,s^{-1}$) is not directed toward anything prominent, but points \emph{away\/} from the Local Void. Some evacuation of voids is expected in standard cosmology, due to the density contrast that pushes matter outward, but the high speeds of galaxies at the edge of the Local Void indicate that it should be very large and very empty, more than usually expected. Consequently, alternative models have been considered. For instance, \citet{dut07} argue that dark energy with a varying equation of state could have higher density where mass density is lower, so that repulsion from voids would be stronger.

Moreover, as discussed by \citet{pee10}, many other properties of galaxies in our neighborhood are not explained by the standard model. Besides the observed extreme emptiness of the Local Void ($\sim10^{-5}$ probability), the presence of large, luminous galaxies in the adjacent low-density regions is in contrast with expectations ($<1$\% probability). These and other observed features seem to indicate the need for a better theory, entailing a mechanism by which matter is more rapidly repelled by voids and gathered into galaxies and structures.

In this paper we want to investigate whether the repulsive gravity scenario proposed by \citet{vil11} as a potential explanation for the cosmic acceleration could also help solve some of these problems. A brief introduction to antimatter gravity is given in Sect.\ 2, while in Sect.\ 3 this view is applied to the dynamics of our extragalactic surroundings. Discussion and conclusions can be found in Sect.\ 4.

\section{Antimatter gravity}

The existence of antigravity, i.e.\ gravitational repulsion between matter and antimatter, has been debated at length during the last several decades, without reaching any firm conclusion. After strong opposition during the second half of last century, the idea of antigravity has experienced a period of renewed interest because of the discovery of the accelerated expansion of the Universe in 1998. In particular, against antigravity in the Fifties--Sixties were proposed some theoretical arguments that seemed to rule out its existence \citep{mor58,sch58,sch59,goo61}, but which were later criticized and questioned \citep[e.g.][]{nie91,cha92,cha93,cha97,haj11a}, and then lost much of their effectiveness. At the same time, several authors have pursued the idea of antigravity, either assuming that antimatter has negative gravitational mass, and thus is self-attractive (e.g.\ \citealp{noy91,ni04,haj10c,haj11c,haj11d}; see also \citealp{cab10}), or that it is even self-repulsive \citep[e.g.][]{noy08,ben12}.

In a recent paper, \citet{vil11} argued that there is no need to change the sign of the gravitational mass of antimatter (which would represent a violation of the weak equivalence principle) to get repulsion between matter and antimatter; but he showed that antigravity appears as a prediction of general relativity, once it is assumed that this theory is CPT invariant and that, consequently, matter is transformed into antimatter by these three joint operations (charge conjugation, parity, and time reversal).

Besides representing the universal symmetry of field theories, the CPT transformation is consistent with the Feynman-Stueckelberg interpretation of antiparticles as particles traveling backwards in time, since this view explains the need of applying T (and P) in addition to C to describe the behavior of antimatter. In other words, a physical system (or a component of a physical system) traveling in the opposite time direction is observed from our time direction through a CPT transformation, and this would be the reason why physical laws are CPT symmetric, since they must describe the behavior of physical systems as observed from either time directions, i.e.\ when they appear to be composed of either matter or antimatter, or both.

In general relativity, the equation of motion for a matter test particle in a matter-generated gravitational field is
\begin{equation}
\label{eq.1}
{{\rm d}^2x^\lambda\over{\rm d}\tau^2}=-\Gamma^{\lambda}_{\mu\nu}{{\rm d}x^\mu\over{\rm d}\tau}{{\rm d}x^\nu\over{\rm d}\tau}\,.
\end{equation}
If we CPT-transform all the four elements in Eq.~(\ref{eq.1}), we get an identical equation describing the motion of an antimatter test particle in an antimatter-generated gravity field, since all the four changes of sign cancel one another. Thus, this CPT symmetry ensures the same self-attractive gravitational behavior for both matter and antimatter. But, if we transform only one of the two components, either the field $\Gamma^{\lambda}_{\mu\nu}$ or the particle (represented by the remaining three elements), we get a change of sign that converts the original gravitational attraction into repulsion, so that matter and antimatter repel each other.

The equation for a massless particle (e.g.\ a photon) is formally equal to Eq.~(\ref{eq.1}), except for the parameter $\tau$, which can no longer be taken as the proper time, being ${\rm d}\tau=0$, but it will be another parameter describing the world line. Therefore, a (retarded) photon will be repelled by an antimatter gravity field, as well as a CPT-transformed photon, i.e.\ an advanced photon, will be repelled by matter\footnote{As a consequence, the energy of a retarded-advanced photon pair in a gravitational field would be conserved, thus invalidating the \citet{mor58} argument against antigravity.}. This may provide a test for the theory of antigravity: the presence of antimatter in cosmic voids suggested by \citet{vil11,vil12a} might be revealed by its gravitational effect on the radiation coming from background sources, in a sort of antigravitational lensing.

In the Newtonian approximation, the law of gravity extended to antimatter becomes $F(r)=\mp GmM/r^2$, where the minus sign refers to the gravitational self-attraction of both matter and antimatter, while the plus sign indicates the gravitational repulsion between matter and antimatter.

The change of sign in the CPT-transformed (i.e.\ antimatter-generated) gravitational field comes from the inversion of the derivatives of the metric tensor in the expression of $\Gamma^{\lambda}_{\mu\nu}$. Thus, while the metric tensor, i.e.\ the potential, does not change sign and remains the same for matter and antimatter, the sign change appears only in the field. This means that in the above generalized Newton law, the plus sign indicating repulsion does not formally come from an inverted potential, but from its inverted gradient. Indeed, the potential $\phi$ is usually defined as $-\nabla\phi={\rm\bf g}$, so that, being the attractive field $g(r)=-GM/r^2$, the potential is $\phi(r)=-GM/r$, and, with the same potential, the repulsive field comes from its inverted gradient: $\bar{g}(r)=GM/r^2$. This formally correct view is actually not intuitive, so that it is preferable to reverse the potential and to get the field according to the usual $-\nabla\phi$ definition. The repulsive potential is thus $\bar{\phi}(r)=GM/r$, and the gravitational picture of the cosmic alternance of matter and antimatter suggested in \citet{vil11} can be represented as an alternation of potential wells and hills. Of course, from the antimatter point of view this situation would be seen upside down.

\section{Dynamics of the Local Sheet}

Our peculiar motion with respect to the general expansion of the Universe can be inferred by the observed dipole anisotropy in the cosmic microwave background \citep[e.g.][]{fix96}. This motion has well-known local contributions, including the orbital velocity of the Sun in our Galaxy and the attraction of the latter towards M31. Once these components are taken into account, it is found that the Local Sheet has still a very large peculiar velocity of $631\rm\,km\,s^{-1}$, which can be decomposed into three quasi-orthogonal components \citep{tul08}. One of them is due to the well-known attraction towards the Virgo Cluster and its dense surroundings \citep[e.g.][]{ton81,aar82,hof82,tul84,ton00}. The second, largest component has a less clear origin: it is ascribed to gravitational pull on scales larger than $3000\rm\,km\,s^{-1}$ nearly in the direction of the Centaurus Cluster, but the identification and relative importance of the ``great attractors" is still debated \citep[e.g.][]{sha84,lyn88,sca89,ray89,koc06,erd06}.

The last component corresponds to what is historically known as the ``local velocity anomaly" \citep[e.g.][]{fab88,tul88,tul92}. It is large, $259\rm\,km\,s^{-1}$, and is anomalous because is not directed towards any important structure. It can not be due to a simple gravitational pull, otherwise also the Leo Spur galaxies, which would be in between, should be affected by it, while they appear to be at rest with respect to this motion. Thus, the cause of this component must be on the other side, where the Local Sheet borders the Local Void. Moreover, this velocity vector points just in the opposite direction with respect to an approximate center of the void, and is orthogonal to the disk-like structure of the Local Sheet. Whatever the cause of the anomalous motion, it must affect the Local Sheet more than the rest of our neighborhood (which is indeed more distant from the void), and also seems to have the property to squash the closest structures, besides giving them high velocities.

These and several other features (see Sects.\ 1 and 4), such as the extreme emptiness of the Local Void, do not find a satisfactory explanation within the standard model, and seem to require the intervention of a repulsive force. In particular, the flattened shape of the Local Sheet looks just like the signature of a repulsive gravitational field. Indeed, while the effect of tidal forces in an attractive gravity field is to stretch structures along the radial direction, since the side closer to the field source is more attracted than the side farther away, the tidal effect of a repulsive field is to squash them, the near side being more pushed than the far one: ${\rm d}F/{\rm d}r=-2GmM/r^3$.

Under some simple assumptions in linear perturbation theory, the peculiar velocity field $\rm{\bf V}_{pec}$ of an ideal pressureless fluid is related to the gravitational acceleration $\rm\bf g$ and the expansion time $t$ by the simple formula \citep[see e.g.][]{pee93,pee11} ${\rm{\bf V}_{pec}}={\rm\bf g}\,t$. We can check whether this equation can represent a reasonable approximation for the peculiar velocity of the Local Sheet with respect to the gravitational attraction by the Virgo Cluster. Rewriting it as
\begin{equation}
\label{eq.2}
\frac{V_{\rm pec}}{\rm km\,s^{-1}}=4.4\times10^{-12}\frac{M}{M_\odot}\frac{\rm Mpc^2}{d^2}\frac{t}{\rm Gyr}\,,
\end{equation}
with $t=t_0=13.75\rm\,Gyr$ \citep{jar11}, $d=17\rm\,Mpc$ and $V_{\rm pec}=185\rm\,km\,s^{-1}$ \citep{tul08}, we get for the mass of the Virgo Cluster and surroundings $M_{\rm Vir}=8.8\times10^{14}\,M_\odot$. This value is in good agreement with previous estimates \citep[e.g.][]{tul84,moh05}, so that we may be confident in Eq.~(\ref{eq.2}), at least at the scale of our supercluster.

Assuming that Eq.~(\ref{eq.2}) holds also for repulsive acceleration, we can try a rough estimate of the mass of antimatter (see Sect.\ 2), possibly located around the center of the Local Void, needed to push the Local Sheet at the observed velocity of $259\rm\,km\,s^{-1}$. The expected distance of the Local Sheet from the center of the void may be assumed in the range $25\pm5\rm\,Mpc$ \citep[e.g.][]{tul08}, which gives $M_{\rm LV}=2.7^{+1.2}_{-1.0}\times10^{15}\,M_\odot$. However, it would be strange that this repulsive interaction affected only the Local Sheet, leaving the rest of our neighborhood unaffected and at rest with respect to the Local Void. Indeed, what is measured as a peculiar velocity of $259\rm\,km\,s^{-1}$ is actually the peculiar velocity of the Local Sheet relatively to the rest of the Local Supercluster, i.e.\ mainly to the Virgo Cluster, and in particular to the Leo Spur galaxies, which are both farther than the Local Sheet from the approximate center of the Local Void by about 7--$8\rm\,Mpc$. Thus, for a more significant estimate of the mass in the Local Void, we take the difference of Eq.~(\ref{eq.2}) between the distance of the Local Sheet (again assumed as $25\pm5\rm\,Mpc$) and the Leo-Spur/Virgo radial distance from the void center (correspondingly estimated as $32.5\pm4.5\rm\,Mpc$), which yields $M_{\rm LV}=6.6^{+4.6}_{-3.1}\times10^{15}\,M_\odot$, i.e.\ significantly larger than the former result, but of the same order of magnitude. These estimates come from the simplistic assumption that the various components are affected, in the radial direction from the void center, only by the dominant repulsive interaction with an antimatter mass concentrated close to the assumed void center, and disregarding other possible contributions in the same direction. Thus, we are not claiming that these values must be considered as extremely realistic, and we can not even choose between the two results, but we can say that an antimatter mass of $\sim5\times10^{15}\,M_\odot$, within a factor of 2, can easily account for the local velocity anomaly. This value is comfortably comparable to the mass of a medium-size supercluster, which means that we would have found the location for antimatter in a matter-antimatter symmetric Universe.

Another argument in favor of the above picture and mass estimate is the following one. We can calculate the total energy per unit mass, given by $V^2/2+GM/d$ (where $V=H_0d+V_{\rm pec}$ is the total velocity with respect to the void center) at the Local Sheet and Leo-Spur/Virgo distances. In the intermediate case seen above, with $M_{\rm LV}=6.6\times10^{15}\,M_\odot$, $d_{\rm LS}=25\rm\,Mpc$ and $V_{\rm pec}^{\rm LS}=634\rm\,km\,s^{-1}$ for the Local Sheet, $d_{\rm Leo}=32.5\rm\,Mpc$ and $V_{\rm pec}^{\rm Leo}=375\rm\,km\,s^{-1}$ for Leo-Spur/Virgo\footnote{$V_{\rm pec}^{\rm LS}$ and $V_{\rm pec}^{\rm Leo}$ are calculated from Eq.~(\ref{eq.2}), and have the correct difference of $259\rm\,km\,s^{-1}$ between them.}, and adopting $H_0=74\rm\,km\,s^{-1}\,Mpc^{-1}$ from \citet{tul08}, we obtain $4.2\times10^6$ and $4.7\times10^6\rm\,km^2\,s^{-2}$ for the respective total specific energies. We then find a small difference of about 10\%, which could also be due to the various uncertainties affecting both measures and model assumptions, or could come from slightly different initial conditions for the differently located systems. This approximate energy conservation at different distances from the antigravity center confirms that the dynamics in the radial direction is actually dominated by this repulsive interaction, and supports our scenario.

If we go back in time conserving mechanical energy, we find $V=0$ at $d=6.7\rm\,Mpc$ for the Local Sheet, and at $d=6.0\rm\,Mpc$ for Leo Spur, whereas one could expect much smaller distances, approaching a singularity. This might mean that our data are too imprecise to allow a similar extrapolation in the past, or that our simplistic dynamical assumption of a dominant, point-like-source field can no longer hold at these earlier stages and also surrounding fields should be taken into account, or that some of the energy has been somehow dissipated, or a combination of all these things.

In any case, what we are finding throughout this section is that the repulsive gravity field can account not only for peculiar (anomalous) velocities and acceleration of the Hubble flow, but even for the Hubble flow itself. There seems to be more than enough energy for driving all the Universe expansion, stored in the potential energy, when, going back in time, matter get closer and closer to the peaks of the antimatter potential hills. No need for initial (artificial) velocities provided by an explosive ``Big Bang", whose energy would not have any identified origin. No need for unknown dark energy to drive the cosmic acceleration, or dark energy is nothing else than the potential energy of gravitational repulsion.

\section{Discussion and conclusions}

There remains the question of why antimatter in voids should not be visible. It seems that something ``dark" must necessarily exist: dark matter, dark energy, and now ``\emph{dark repulsors\/}". On the other hand, were it otherwise, we would have understood everything a long time ago.

Like matter, antimatter is self-attractive, so we can expect that it forms anti-galaxies and anti-stars, which would emit electromagnetic radiation, we should can detect. However, as pointed out by \citet[][see also Sect.\ 2]{vil12a}, antimatter, if emitting, should emit advanced radiation, which can be undetectable for at least two reasons. First, it could be intrinsically undetectable, at least with the usual detectors and observing procedures\footnote{Actually there is some confusion about advanced radiation in the literature; for discussions around the issue of advanced radiation in general, see e.g.\ \citet{dav75}, \citet{cra80}, and references therein.}. Then, even if it were somehow detectable, we know from Sect.\ 2 that advanced photons are repelled by matter, so that the chance of detecting the few that possibly succeeded in reaching us on top of the potential hill of our Galaxy is further greatly reduced. The possibility remains that antimatter emits also retarded radiation (as well as matter might emit advanced radiation). In this case we should see antimatter stars, galaxies and clusters filling the voids, but well separated from matter ones by gravitational repulsion. Yet, as already said, antimatter would be matter traveling backwards in time, so that, taking this concept to the extreme, we would observe its cosmic evolution in reverse, and its current evolutionary stage could be very different from ours. For example, if antimatter were ``created" in a distant future, now it might be collapsed into supermassive black holes, or be in some other non-emitting form, before definitively collapsing, 14 billion years ago. Perhaps to be ``reborn" as matter, and retrace in the opposite direction the time already passed through, but well separated from the ``anti-itself" by gravitational repulsion. A further step would lead us to suggest a Universe ``recurring over time", in a cosmic loop, but such a speculative rumination is beyond the scope of this paper.

In any case, there seem to be more reasons for antimatter invisibility than for visibility, so that we are not surprised not to see anything in cosmic voids. Another point is the different, quasi-spherical symmetry of voids with respect to the squashed/filamentary shape of matter structures. This would support the view of different evolutionary stages, where antimatter black holes close to the void centers produce spherical fields, and the more uniformly distributed matter can only adapt in the potential valleys among the quasi-circular potential hills.

Passing several cosmological tests, the $\Lambda$CDM standard model seems to provide a fairly good description of the expanding Universe\footnote{However, in a recent paper, \citet{ben12} have shown that most of these observational constraints can also be explained by their quite different cosmology, based on the assumption of a matter-antimatter symmetric Universe.}, but at the cost of introducing ad hoc elements that do not have a physical explanation, like the initial ``explosion" causing the cosmic expansion and the dark energy that should accelerate it. As we have seen, antigravity can explain both the expansion and its acceleration, and this theory also has the virtue of solving the long-standing problem of matter-antimatter asymmetry. It might seem that one is replacing something invisible (dark energy) with something else equally invisible (dark repulsors). The difference is that the presence of the latter is theoretically expected if not required, whilst the presence of the former is not requested, unexpected, and embarrassing.

Moreover, some recent observational findings are undermining the standard model (e.g.\ \citealp{wal11} on unexpectedly uniform distribution of dark matter in dwarf galaxies\footnote{See \citet{haj11c} for an alternative interpretation of dark matter based on antigravity.}). In particular, the existence of ``dark flows", i.e.\ large-scale, peculiar bulk motions of galaxy clusters (e.g.\ \citealp{kas10} and references therein), of excess clustering on large scales \citep{tho11}, of anisotropy in the expansion acceleration \citep[e.g.][]{cai12} seem to indicate the need for a distribution of acceleration sources less homogeneous than the uniformly permeating dark energy, thus favoring the discrete distribution in cosmic voids.

Let us return to our extragalactic neighborhood and its anomalies partly mentioned in the Introduction and described in detail in \citet{pee10}. All those observed properties, such as the extreme emptiness of the Local Void, the unexpected presence of large galaxies at its edge, the existence of pure disk galaxies, and the insensitivity of galaxy properties to environments, indicate the existence of a mechanism that can produce a fast evacuation of voids and rapid assembly of matter into structures, which the standard theory based on attractive-only gravity can not provide. Repulsive gravity can be the cause of fast evacuation and, also through the squashing tidal effect, can help a lot to rapidly gather matter into galaxies and structures, especially at the borders of voids.

In conclusion, repulsive gravity between matter and antimatter, which is a natural outcome of general relativity extended to CPT-transformed systems, is an excellent candidate to explain not only the cosmic expansion and its acceleration, but also the anomalous kinematics and several other properties of our extragalactic surroundings, as well as other observed features of the Universe that can not be accounted for by the standard model and its mysterious ingredients.

%%%aggiunto
\makeatletter
\let\clear@thebibliography@page=\relax
\makeatother

%%%\bibliographystyle{spr-mp-nameyear-cnd}
%\bibliographystyle{aa}
%\bibliography{villata}

%\end{document}

\end{document}